# H-MAC: A HYBRID MAC PROTOCOL FOR WIRELESS SENSOR NETWORKS[1]

S. Mehta and K.S. Kwak

UWB Wireless Communications Research Center, Inha University
Incheon, 402-751, Korea
suryanand.m@gmail.com

### *ABSTRACT*

*In this paper, we propose a hybrid medium access control protocol (H-MAC) for wireless sensor networks. It is based on the IEEE 802.11's power saving mechanism (PSM) and slotted aloha, and utilizes multiple slots dynamically to improve performance. Existing MAC protocols for sensor networks reduce energy consumptions by introducing variation in an active/sleep mechanism. But they may not provide energy efficiency in varying traffic conditions as well as they did not address Quality of Service (QoS) issues. H-MAC, the propose MAC protocol maintains energy efficiency as well as QoS issues like latency, throughput, and channel utilization. Our numerical results show that H-MAC has significant improvements in QoS parameters than the existing MAC protocols for sensor networks while consuming comparable amount of energy.*

## 1. INTRODUCTION

Wireless sensor networking is an emerging technology that has a wide range of potential applications including monitoring, medical systems and robotic exploration. Sensor networks normally consists of a large number of densely deployed distributed nodes that organize themselves into a multi-hop wireless networks. The wireless sensor nodes are usually equipped with limited power source. In some application scenarios, replacement of power resource might be not possible. It is for these reasons that researcher are currently focusing on the design of power-aware protocols and schemes for sensor networks. These schemes include power saving hardware design, power saving topology design and power efficient MAC layer protocols, to name a few [1]. Communication in wireless sensor networks is divided into several layers. One of those is the Medium Access Control (MAC) layer. MAC is an important technique that enables the successful operation of the network. MAC protocol tries to avoid collisions so that two interfering nodes do not transmit at the same time. There are some MAC protocols that have been developed for wireless sensor networks. Typical examples include S-MAC and T-MAC [2][3]. The main design goal of the typical MAC protocols is to provide high throughput and QoS. On the other hand, wireless sensor MAC protocol gives higher priority to minimize energy consumption than QoS requirements. Energy gets wasted in traditional MAC layer protocols due to idle listening, collision, protocol overhead, and over-hearing [2].

S-MAC is proposed to improve energy efficiency in wireless sensor networks. It divides the time into large frames. Every frame has two parts: an active part (on time) and a sleeping part. A node turns off its radio during the sleep time to preserve the energy. During the active time, a node can communicate with its neighbors and transmits the queued packets during the sleeping time. Hence, S-MAC saves the unnecessary waste of energy on idle listing. Periodic sleep may result in high latency (for mutihop sensor networks) and low throughput problems [2]. Timeout-MAC protocol is proposed to enhance the performance of S-MAC under variable traffic conditions. It defines the minimum time out timer TA. T-MAC works similar to S-MAC with one difference of time out timer. In T-MAC, active time ends when no activation has occurred for a time out timer TA. Time-out timer may result in early sleeping and low throughput problems [3]. Data gathering MAC falls into duty cycle based MAC protocols category. The main aim of D-MAC is to achieve low latency as well as energy-efficiency. D-MAC is an improved slotted aloha protocol where slots are assigned to the sets of nodes based on data gathering tree (Parent-child topology). In D-MAC, low latency is achieved by assigning subsequent slots to the nodes that are successive in the data transmission path. However, D-MAC works well only for tree based structure and may cause a collision problem [4]. Pattern-MAC is a 'time slotted' protocol. It adaptively determines the sleep-wake up schedules for a node based on its own traffic and the traffic patterns of its neighbours. In P-MAC, a node gets information about the activity in its neighbourhood before hand through patterns. Based on theses patterns, a node can put itself into a long sleep for several time frames when there is no traffic in the network. If there is any activity in the neighbourhood, a node will know this through the patterns and will wake-up when required. Hence, P-MAC saves more energy compared to S-MAC without compromising on the throughput. But it is worthwhile to mention about complexity, collision, and overhead problems in P-MAC [5]. To maximize the battery lifetime, sensor networks MAC protocols implement the variation of active/sleep mechanism. This mechanism trades networks QoS for energy savings. However, the propose H-MAC not only reduce the comparable amount of energy but also gets good QoS such as latency, throughput, and channel utilization. From [2]-[5], we can compare the previous works on different points, and summarize them as shown in table 1.

[1] "A part of this paper was published in IEEE Seoul Section 2008 "Poster Compettition" as poster , Seoul,Korea [7]."

Table 1. Comparison table

| Protocol | S-MAC | T-MAC | D-MAC | P-MAC | H-MAC |
|---|---|---|---|---|---|
| Time-Sync. | Yes | Yes | Yes | Yes | Yes |
| Point to Point | Suitable | Suitable | Not-Suitable | Suitable | Suitable |
| Broadcast | Not-Suitable | Not-Suitable | Not-Suitable | Suitable | Suitable |
| Convergecast | Not-Suitable | Not-Suitable | Suitable | Suitable | Suitable |
| Mobility | Not-Suitable | Not-Suitable | Not-Suitable | May be | May Be |
| Type | CSMA | CSMA | TDMA/S. Aloha | Slotted Aloha | CSMA/S. Aloha |
| Adaptive to change | Ok | Good | Weak | Good | Good |
| Half/Full Duplex | Half | Half | Full | Full | Full |

## 2. H-MAC PROTOCOL

We present a new hybrid MAC protocol- H-MAC, for sensor networks. H-MAC is based on IEEE 802.11's PSM mode and slotted aloha [6]. In H-MAC, time is divided into large frames, every frame has two parts: an active part (on time) and a sleeping part. Active part is like ATIM window in PSM mode and sleeping part is further divided into N slots, where each slot is bit bigger than data frame. Figure 1 shows the comparison between S-MAC and H-MAC time frames.

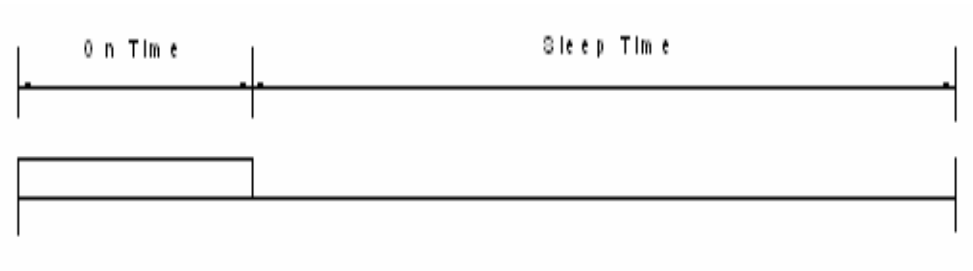

(a) A time frame of S-MAC protocol

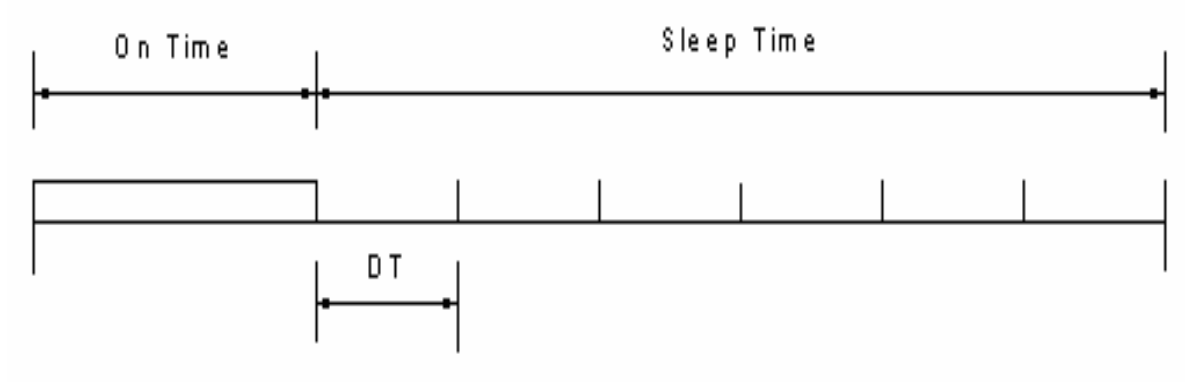

(b) A time frame of H-MAC protocol

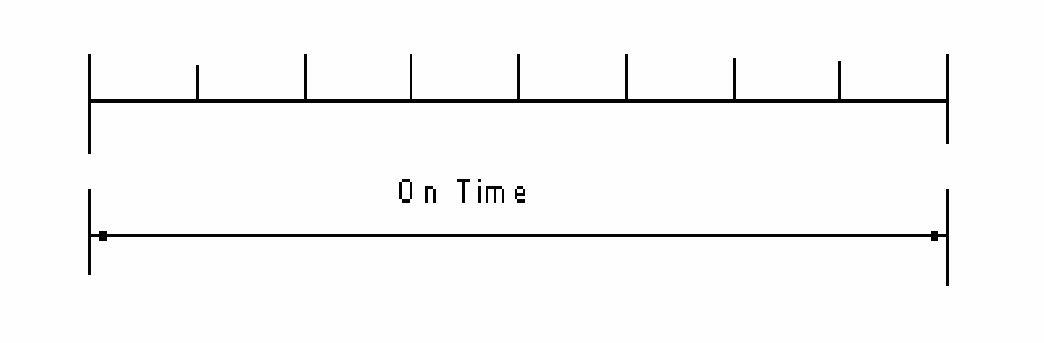

(c) On time frame of H-MAC protocol

Figure 1. Time frame of S-MAC and H-MAC protocols

The nodes that have packets to transmit negotiate slots with the destination nodes during active time and transmit/receive the data packets in pre-negotiated slots during sleep time. If the nodes don't have to transmit or receive any data packets go to sleep during the sleep-time slots. Figure 2 illustrates the working of H-MAC.

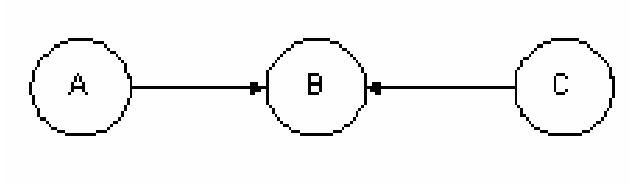

(a) An example topology

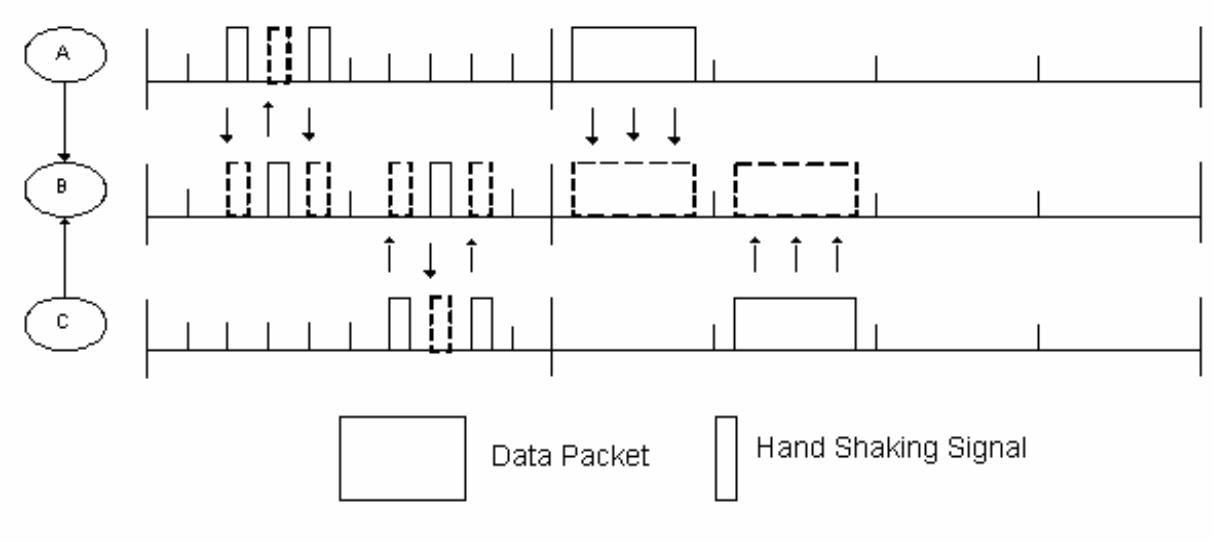

(b) Data transmission in H-MAC protocol

Figure 2. Working of H-MAC protocol

If node A has buffered packets destined for node B, it will notify node B by sending ATIM packet. Node A includes its preferable slot(s) list in the ATIM packet. Node B, upon receiving the ATIM packet, select slot(s) based on sender's list and its own list. The receiver's list has higher priority in selecting the slot(s). After Node B selects a slot(s), it includes the slot information in the ATIM-ACK packet and sends it to node A. When node A receives the ATIM-ACK packet, it sees if it can also select the slot(s) specified in the ATIM-ACK. If node A selects the slot(s) specified in the ATIM-ACK, node A sends an ATIM-RES (ATIM-Reservation) packet to the node B, with node A's selected slot(s) specified in the packet. The ATIM-RES is a new type of packet used in our MAC scheme, which is not in IEEE 802.11 PSM. The ATIM-RES packet notifies the nodes in the vicinity of node A which slot(s) node A is going to use, so that the neighbouring nodes can use this information to update their list. Similarly, the ATIM-ACK packet notifies the nodes in the vicinity of node B. After the ATIM (On time) time, node A and node B will transfer the data packet(s) in selected slot(s).

## 3. NUMERICAL RESULTS

In this section, we show latency and throughput analysis of S-MAC and H-MAC protocols. As S-MAC is widely accepted and popular sensor networks MAC protocol, we choose S-MAC for comparison with H-MAC [2, 7]. For our calculation we consider 10 hops liner topology as shown in figure 3.

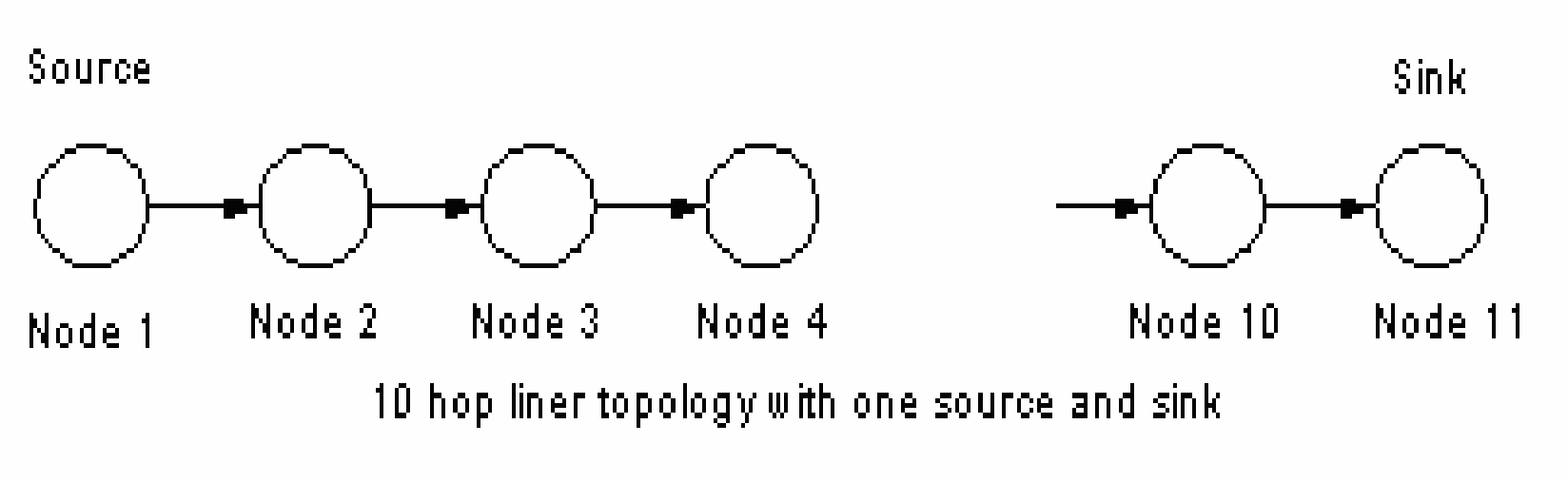

Figure 3. 10 hops liner topology

And we also consider 20 slots per cycle, where 18 slots are used for data transfer and 2 slots for handshaking signals.

### 3.1. Latency Analysis:

IEEE 802.11 Protocol [8]:

The entire latency over N hops is given by

$$D(N) = \sum_{n=1}^{N} (t_{cs,n} + t_{tx}). \quad (1)$$

Where, N is the number of hops. $t_{cs,n}$ and $t_{tx}$ represents backoff and transmission delay respectively. And $n$ represents the current hop value, average latency over N hops is given by

$$E[D(N)] = N(t_{cs,n} + t_{tx}) \quad (2)$$

S-MAC Protocol[2]:

In S-MAC, a complete cycle is denoted by $T_f$ and has two parts: an active part and a sleep part. Listen/active time is fixed and set to 10% of $T_f$ (10% duty cycle). The delay at hop $n$ is given by

[2] As adaptive listening mode increases the energy consumption; we are not considering this mode of S-MAC. However, adaptive listing mode reduces the latency at the cost of energy-efficiency.

$$D_n = t_{s,n} + t_{cs,n} + t_{tx} \tag{3}$$

Where, $T_f >> t_{tx}$ and $t_{s,n}$ is the sleep delay. In S-MAC without adaptive listening, contention only starts at the beginning of each frame. This is given by

$$T_f = t_{cs,n-1} + t_{tx} + t_{s,n} \tag{4}$$

So the sleep delay at hop n is given by

$$t_{s,n} = T_f - (t_{cs,n-1} + t_{tx}). \tag{5}$$

Substituting (5) in to (3)

$$D_n = T_f + t_{cs,n} - t_{cs,n-1} \tag{6}$$

A packet can be generated on the source node at any time within a frame, so the sleep delay on the first hop, $t_{s,1}$, is a random variable whose value lies in $(0, T_f)$. Suppose $t_{s,1}$ is uniformly distributed in $(0, T_f)$. Its mean value is $T_f/2$. Combining it with (6), the overall delay of a packet over N hops is given by

$$\begin{aligned} D(N) &= D_1 + \sum_{n=2}^{N} D_n \\ &= t_{s,1} + t_{cs,1} + t_{tx} + \sum_{n=2}^{N} \left( T_f + t_{cs,n} - t_{cs,n-1} \right) \\ &= t_{s,1} + (N-1)T_f + t_{cs,N} + t_{tx} \end{aligned} \tag{7}$$

The average latency of S-MAC without adaptive listen over N hops is given by

$$\begin{aligned} E[D(N)] &= E[t_{s,1} + (N-1)T_f + T_{cs,N} + t_{tx}] \\ &= T_f/2 + (N-1)T_f + t_{cs} + t_{tx} \\ &= NT_f - T_f/2 + t_{cs} + t_{tx}. \end{aligned} \tag{8}$$

From the (8) we can observe that the multihop latency linearly increases with the number of hops in S-MAC when each node strictly follows its sleep schedule(s).

H-MAC Protocol:

H-MAC is similar to S-MAC with only one difference of slotted sleep time.

In H-MAC, time frame $T_{f-X}$ is given by

$$T_{f-x} = t_{active} + Ct_{s,n} \tag{9}$$

Where C is the number of equal length slots and $t_{active}$ is the listen time same as S-MAC (10% duty cycle). During the $t_{active}$ time H-MAC can reserve the slots for $n_r$ hops within the same $T_{f-x}$, so the delay for 1 hop transmission over $n_r$ hops is given by

$$D_1 \approx T_{f-x}/n_r \tag{10}$$

From the (10), we can also calculate the delay for N hops as

$$D(N) \approx \sum_{n=1}^{N} D_1 \qquad (11)$$

From (10) and (11), we can calculate the average latency of H-MAC as

$$\begin{aligned} E[D(N)] &\approx E[\sum_{n=1}^{N} D_1] \\ &\approx N(T_{f-x}/n_r) \\ &\approx \lceil N(T_{f-x}/n_r) \rceil \end{aligned} \qquad (12)$$

Where $\lceil * \rceil$ define the largest integer value which is equal to *. From (12) we can observe that the multihop latency linearly increases with the number of hops in H-MAC as in S-MAC. However, the slop of the line changes to $T_{f-x}/n_r$ .

## 3.2. Throughput Analysis:

Here, packet length is fixed and represented by $t_p$. Sleep time is represented by $t_{Sleep}$ and equivalent to $Ct_s$. Actual data transmission take place only during the sleep time and $t_{Sleep} >> t_p$. Hence, the throughput is given by

$$Th = \frac{t_{sleep}}{t_{active} + t_{sleep}} \qquad (13)$$

S-MAC Protocol:

In S-MAC, a node can communicate $n_p$ packets to only one node within a frame time, so the throughput is given by

$$Th_s = \frac{n_p}{t_{active} + t_{sleep}} \qquad (14)$$

H-MAC Protocol:

In H-MAC, a node can communicate $n_p$ packets to maximum $n_m$ nodes within a frame time. So the throughput is given by

$$\begin{aligned} Th_x &= \frac{Ct_s}{t_{active} + Ct_s} \\ &= \frac{n_p n_m}{t_{active} + Ct_s} \end{aligned} \qquad (15)$$

from (14) and (15) it is clear that H-MAC gives better throughput condition compared to S-MAC.

Now we present analysis to find the ratio of successful transmitted messages[3] [9]. Here, active time is fixed and equivalent to $S_m$ slots. There are $N$ nodes to compete for medium/slots. A node can transmit only one request.

Let $n$ be the number of nodes tries to get the same mini-slot among $N$ nodes. The request messages are uniformly distributed in an active time. The probability that $n$ nodes are in a slot is given by binomial distribution as follows

$$P[X = n] = \binom{N}{n}\left(\frac{1}{S_m}\right)^n \left(1-\frac{1}{S_m}\right)^{N-n} \qquad (16)$$

The above binomial distribution also applies to $S_m$ slots, thus the expected value of the number of slots with n nodes in a slot is given by

$$E[X = n] = S_m \binom{N}{n}\left(\frac{1}{S_m}\right)^n \left(1-\frac{1}{S_m}\right)^{N-n} \qquad (17)$$

$C_n$ represents the number of slots being filled with exactly $n$ nodes. So the average number of collided messages is given by

$$\begin{aligned}\sigma &= \sum_{n=2}^{N}\sum_{c_n=1}^{S_m} nP[X = C_n]C_n = \sum_{n=2}^{N} nE[X = C_n] \\ &= \sum_{n=2}^{N} nS_m \binom{N}{n}\left(\frac{1}{S_m}\right)^n \left(1-\frac{1}{S_m}\right)^{N-n} \qquad (18) \\ &= N - N\left(1-\frac{1}{S_m}\right)^{N-1}\end{aligned}$$

from (17) and (18) we can calculate the ratio of the number of successfully transmitted request messages and the total number of transmitted request messages. The ratio is given by

$$Ratio = \frac{N-\sigma}{N} = \left(1-\frac{1}{S_m}\right)^{N-1} \qquad (19)$$

Figure 4 shows the throughput of the network while varying the number of packets transmitted from sink to source. After 8 packets H-MAC's throughput reduces to S-MAC throughput, as H-MAC can't transmit the packets simultaneously above 8 packets and data packets take more cycles to reach sink[4]. Figure 5 shows the energy consumption of the network, S-MAC and H-MAC consumes the same amount energy. Figure 6 shows the latency performance of S-MAC and H-MAC. H-MAC performs notably well compared to S-MAC, as it can transmit the data simultaneously to other nodes. Figure 7 and 8 shows the ratio of successful transmitted message. First result we obtain by varying the number of neighbouring nodes, and keeping cycle size of 20 slots. Similarly, second result we obtain by varying the number of slots in given cycle, and keeping the neighbouring nodes constant.

---

3 Only for H-MAC protocol.

4 Here, we simplified our result for simplicity; normally H-MAC will perform better than S-MAC protocol

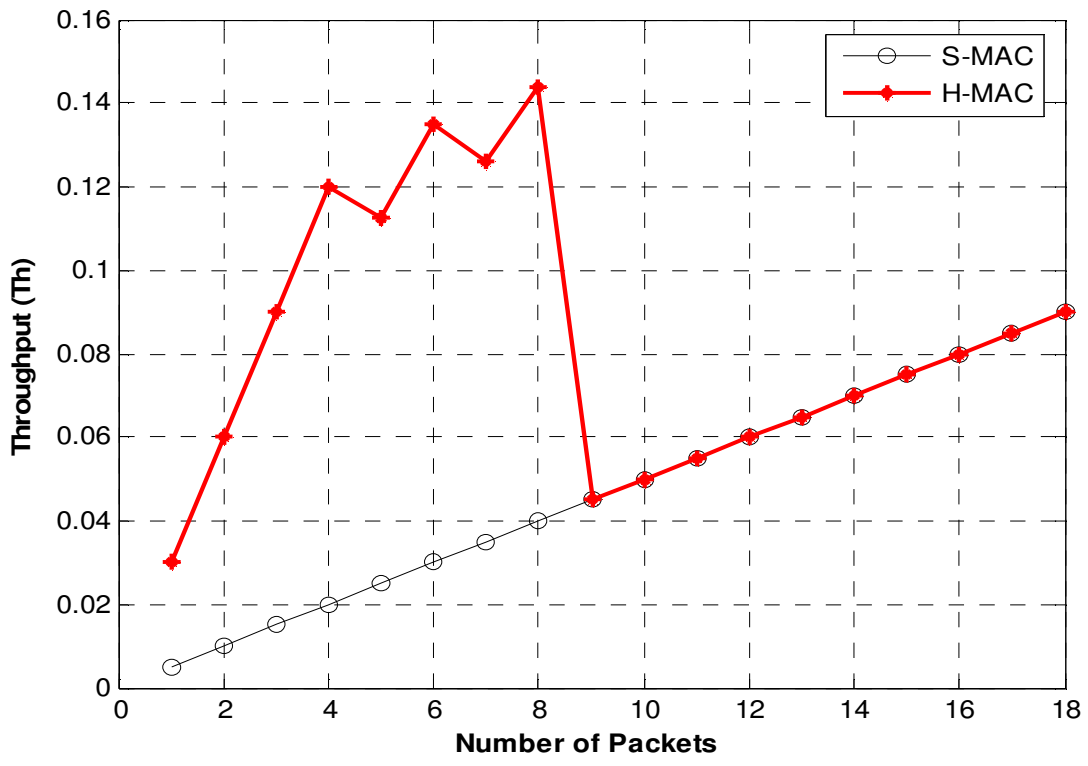

Figure 4. Throughput of the network

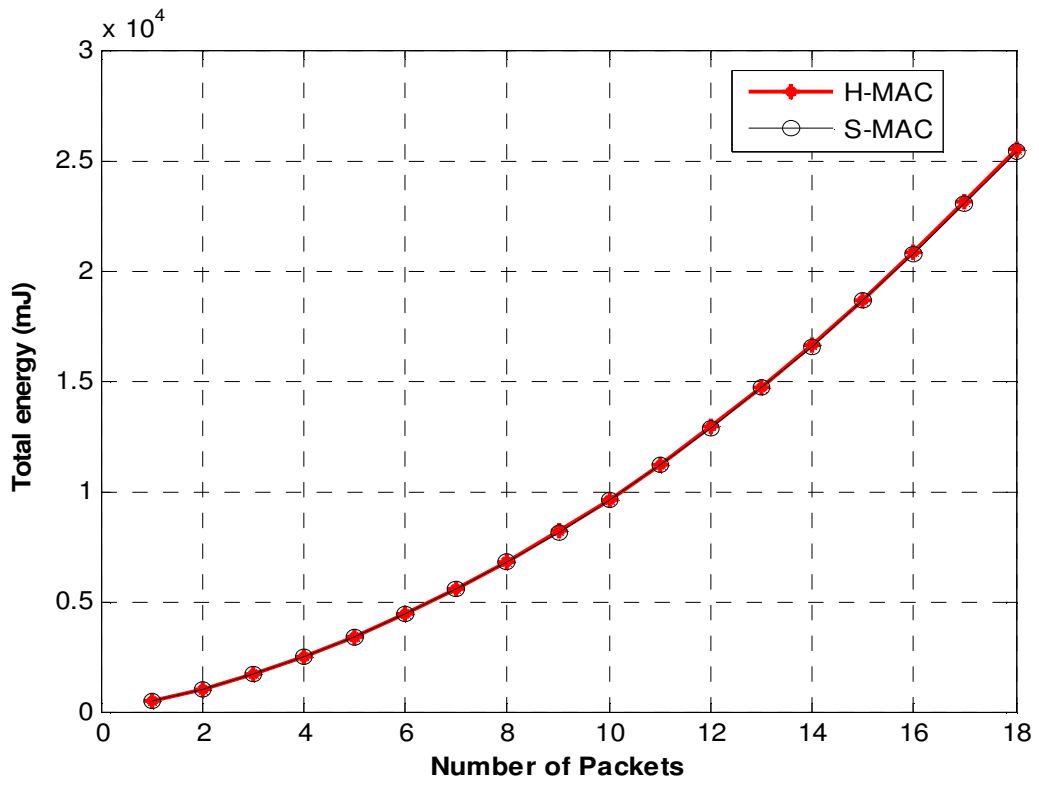

Figure 5. Total energy consumption

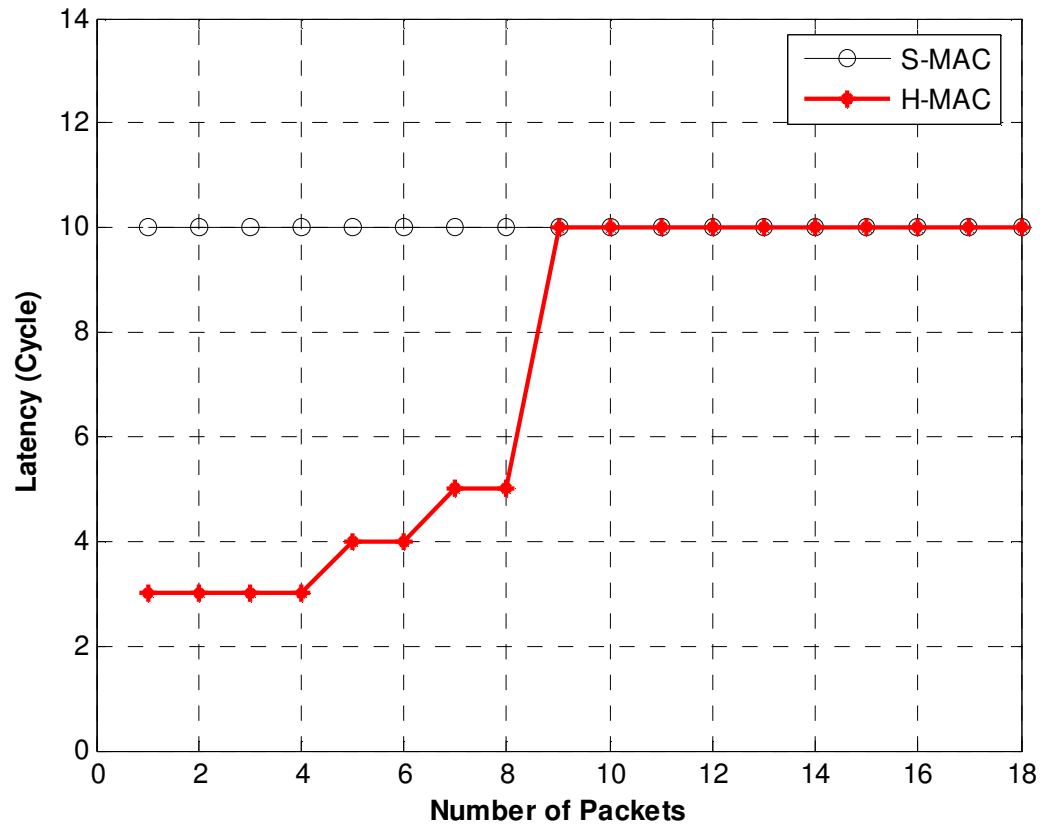

Figure 6. Latency performance

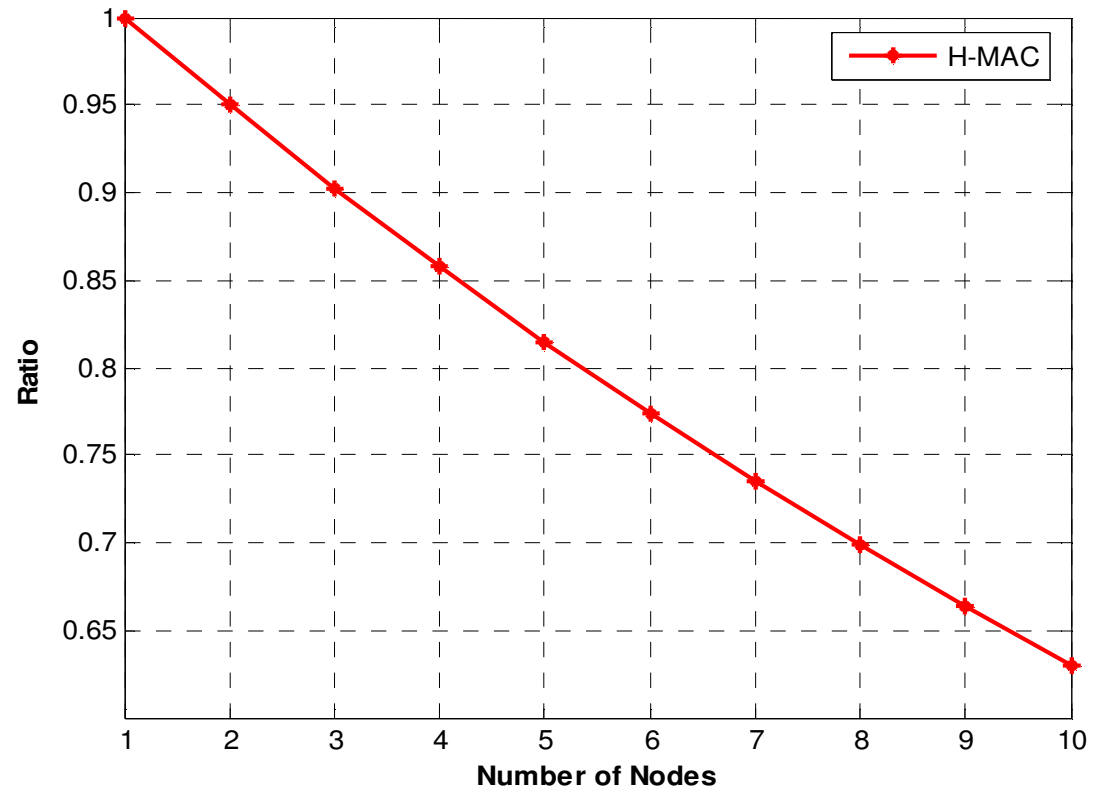

Figure 7. Ratio of successful transmitted messages

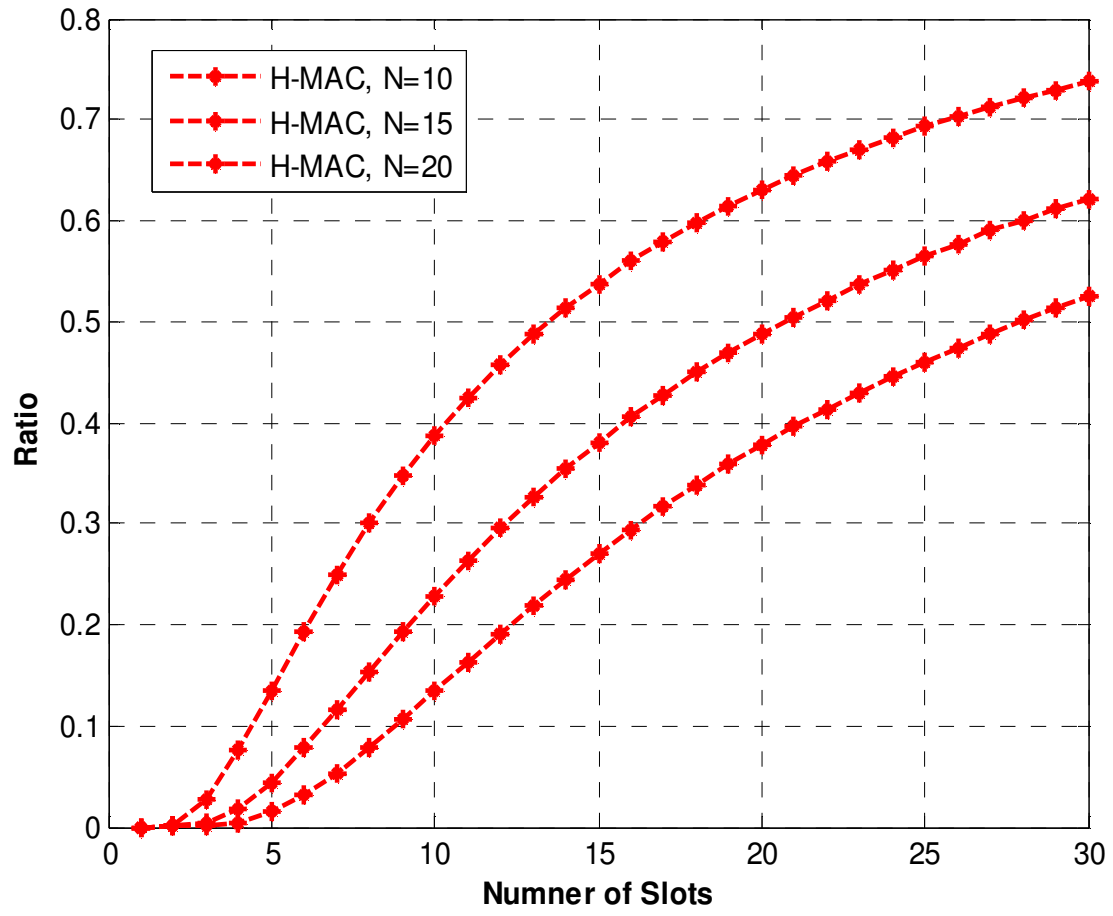

Figure 8. Ratio of successful transmitted messages

## 4. CONCLUSIONS

In this paper, we present the H-MAC protocol, a hybrid MAC protocol based on IEEE 802.11's PSM mode and slotted aloha, and utilizes multiple slots dynamically to improve performance. We also present the numerical results of H-MAC, and show that H-MAC has significant improvements in QoS parameters than the existing MAC protocols for sensor networks while consuming comparable amount of energy.

**Authors**

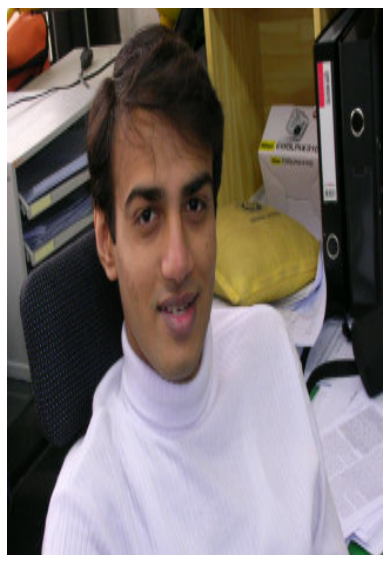

**S.Mehta** received the B.E. and M.S degrees both in Electronics Engineering from Mumbai University, Mumbai, India, and Ajou University, Korea in 2002 and 2005, respectively. He is currently pursuing the Ph.D. degree in Telecommunication engineering from the Inha University, Korea. His research interests are in the area performance analysis of wireless networks and RFID systems.

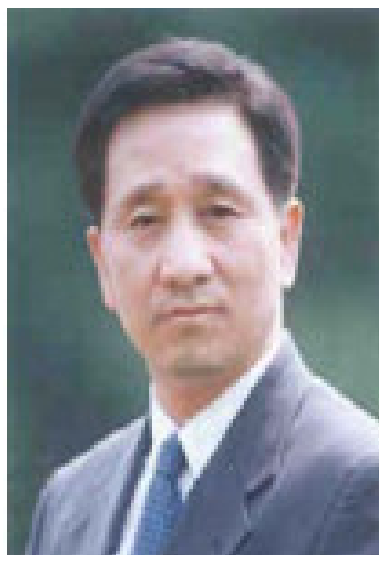

**K. S. Kwak** received the B.S. degree from Inha University, Korea in 1977, and the M.S. degree from the University of Southern California in 1981 and the Ph.D. degree from the University of California at San Diego in 1988, respectively. From 1988 to 1989 he was a Member of Technical Staff at Hughes Network Systems, San Diego, California. From 1989 to 1990 he was with the IBM Network Analysis Center at Research Triangle Park, North Carolina. Since then he has been with the School of Information and Communication, Inha University, Korea as a professor. He had been the chairman of the School of Electrical and Computer Engineering from 1999 to 2000 and the dean of the Graduate School of Information Technology and Telecommunications from 2001 to 2002 at Inha University, Inchon, Korea. He is the current director of Advanced IT Research Center of Inha University, and UWB Wireless Communications Research Center, a key government IT research center, Korea. He has been the Korean Institute of Communication Sciences (KICS)'s president of 2006 year term. In 1993, he received Engineering College Young Investigator Achievement Award from Inha University, and a distinguished service medal from the Institute of Electronics Engineers of Korea (IEEK). In 1996 and 1999, he received distinguished service medals from the KICS. He received the Inha University Engineering Paper Award and the LG Paper Award in 1998, and Motorola Paper Award in 2000. His research interests include multiple access communication systems, mobile communication systems, UWB radio systems and ad-hoc networks, high-performance wireless Internet.